\documentclass[prd,twocolumn,showpacs,preprintnumbers,amsmath,amssymb,superscriptaddress,floatfix,nofootinbib,article]{revtex4}
\usepackage{txfonts}
\usepackage{graphicx}
\usepackage{amsmath}
\usepackage{amsfonts}
\usepackage{amssymb}
\usepackage{color}
\usepackage{subfigure}
\usepackage{multirow}
\usepackage[colorlinks, citecolor=blue,anchorcolor=red,menucolor=red,linkcolor=red,filecolor=red,runcolor=red,urlcolor=blue,frenchlinks=red]{hyperref}

\begin{document}
 \title{Roles of the scalar $f_0(500)$ and $f_0(980)$ in the process $D^0\to \pi^0\pi^0 \bar{K}^0$}

\author{Xiao-Hui Zhang}
\affiliation{School of Physics, Zhengzhou University, Zhengzhou 450001, China}

 \author{Han Zhang}
\affiliation{School of Physics, Zhengzhou University, Zhengzhou 450001, China}	

\author{Bai-Cian Ke}
\affiliation{School of Physics, Zhengzhou University, Zhengzhou 450001, China}

\author{Li-Juan Liu}\email{liulijuan@zzu.edu.cn}
\affiliation{School of Physics, Zhengzhou University, Zhengzhou 450001, China}

\author{De-Min Li}\email{lidm@zzu.edu.cn}
\affiliation{School of Physics, Zhengzhou University, Zhengzhou 450001, China}

  \author{En Wang}\email{wangen@zzu.edu.cn}
\affiliation{School of Physics, Zhengzhou University, Zhengzhou 450001, China}

\begin{abstract}
Motivated by the near-threshold enhancement and the dip structure around 1~GeV in the $\pi^0\pi^0$ invariant mass distribution of the process $D^0\to \pi^0\pi^0\bar{K}^0$ observed by the CLEO Collaboration,  
we have investigated this process by taking into account the contribution from the $S$-wave pseudoscalar meson-pseudoscalar meson interactions within the chiral unitary approach, and also the one from  the intermediate resonance $\bar{K}^{*}(892)$. 
Our results are in good agreement with the CLEO measurements, which implies that, the near-threshold enhancement near the $\pi^0\pi^0$ threshold is mainly due to the contributions from the scalar meson $f_0(500)$ and the intermediate $\bar{K}^*$, and 
the cusp structure around 1~GeV in the $\pi^0\pi^0$ invariant mass distribution should be associated with the scalar meson $f_0(980)$.  	
\end{abstract}
 
\maketitle
 	
 \section{Introduction}
 \label{sec1}
 	
 In the classical quark model, hadrons are categorized into mesons and baryons, where mesons are composed of a quark-antiquark pair, and baryons are composed of three quarks~\cite{Gell-Mann:1964ewy,Zweig:1964ruk}. Although most of the mesons could be well described within the classical quark models, some mesons have exotic properties which are difficult to be explained. For instance, the light scalar mesons $a_0(980)$, $f_0(980)$, and $f_0(500)$ should have more complicate components, and attract many theoretical studies~\cite{Close:2002zu}. However, there are still different explanations for their structure, such as compact tetraquark, molecular state, and the mixing of different components~\cite{Baru:2003qq,tHooft:2008rus,Klempt:2007cp,Kaiser:1998fi,Oller:1997ti,Achasov:2020fee,Morgan:1974cm,Jaffe:1976ig,Deng:2012wj}.

The hadronic decays of the charmed hadrons could provide valuable insights for understanding the short distance weak interaction and the  long-distance Quantum Chromodynamics (QCD) interaction~\cite{Wang:2021ail,Lyu:2024qgc,Duan:2024led,Li:2024rqb,Zhang:2024jby}. Especially, the final state interaction are the key ingredients in the production of the light scalar mesons, i.e. $f_0(500)$, $f_0(980)$, and $a_0(980)$, which are of particular interest for exploring the nature of the light scalar mesons~\cite{Liang:2016hmr,Debastiani:2016ayp,Oset:2016lyh,Wang:2021naf,Wang:2022nac,Feng:2020jvp,Wang:2020pem}.

 Recently, BESIII, {\it BABAR}, and Belle/Belle II have accumulated a lot of experimental data about the multi-body decays of the charmed hadrons, which provides an important platform to investigate these light scalar mesons~\cite{Wang:2021kka,Liang:2014ama,Xie:2014gla,BaBar:2010nhz,Ling:2021qzl,Ding:2024lqk,Ding:2023eps}. For instance, the BESIII and {\it BABAR} Collaborations have performed the amplitude analysis of the process $D_s^+\to K^+K^-\pi^+$, and observed the near-threshold enhancement in the $K^+K^-$ invariant mass distribution~\cite{BESIII:2020ctr,BaBar:2010wqe}. In Ref.~\cite{Wang:2021naf}, the authors have investigated this process by taking into account the $S$-wave pseudoscalar-pseudoscalar interactions within the chiral unitary approach, and concluded that scalar meson $f_0(980)$ gives the dominant contribution for the near-threshold enhancement in the $K^+K^-$ invariant mass distribution.  In addition, the authors of Ref.~\cite{Zhang:2022xpf} have studied the decay $D^+_s\to \pi^+\pi^0\pi^0$, and concluded that the cusp structure around 1~GeV in the $\pi^0\pi^0$ invariant mass distribution could be associated with the scalar meson $f_0(980)$, which favors the no-$q\bar{q}$ nature of $f_0(980)$.

In 2011, the CLEO Collaboration has conducted a Dalitz plot analysis on $D^0 \to K_S^0 \pi^0\pi^0$ using data set of 818 pb$^{-1}$ of $e^+e^-$ collisions accumulated at $\sqrt{s}=3.77$~GeV~\cite{CLEO:2011cnt}, and concluded that combined $\pi^0\pi^0$ $S$-wave contributes $(28.9\pm6.3\pm 3.1)\%$ of the total decay rate, while the $D^0\to \bar{K}^{*0} \pi^0$  contributes $(65.6\pm 5.3\pm 2.5)\%$. In addition, the decay branching fraction of the $D^0 \to \bar{K}^0 \pi^0\pi^0$ was also measured to be $(1.058\pm0.038\pm0.073)\%$~\cite{CLEO:2011cnt}. 
In the $\pi^0\pi^0$ invariant mass distribution, a cusp structure followed by a dip appears around 1~GeV, which may be associated with the scalar meson $f_0(980)$, and an enhancement was also found near the $\pi^0\pi^0$ threshold. Meanwhile, one could find a clear peak of $\bar{K}^*(892)$ in the $\pi^0\bar{K}$ invariant mass distribution. Due to the conservation of the charge parity, there is no contribution from the intermediate vector meson $\rho$ in the process  $D^0\to \bar{K}^0\pi^0\pi^0$,  thus this mode could provide a more clear way to investigate the $S$-wave $\pi\pi$ substructure than the process $D^0\to K^0_s\pi^+\pi^-$~\cite{BaBar:2005rek,CLEO:2002uvu}.

  In this work, we will investigate the process $D^0\to \pi^0\pi^0
  \bar{K}^0$ by considering the scalar $f_0(980)$ generated from the $S$-wave pseudoscalar-pseudoscalar meson interaction within the chiral unitary approach. In addition, we also consider the contribution from the intermediate vector meson $\bar{K}^*(892)$.
   
 This paper is organized as follows. In Sec.~\ref{sec2}, we will present the mechanism for the process $D^0\to \pi^0\pi^0\bar{K}^0$, and our results and discussions will be shown in Sec.~\ref{sec.3}, followed by a short summary in the last section.
   
 \section{formalism}
 \label{sec2}
 	
 In Sec.~\ref{sec2a}, we first present the mechanism of the Cabibbo-favored process $ D^0 \to \bar{K}^0 \pi^0\pi^0$ via the $S$-wave pseudoscalar meson-pseudoscalar meson interactions, which could dynamically generated the scalar mesons $f_0(500)$ and $f_0(980)$. Meanwhile, this process could also happen via the intermediate vector meson $\bar{K}^*(892)$, and the corresponding formalism and the double differential decay width are given in Sec.~\ref{sec.2b}.
 	
 \subsection{$S$-wave pseudoscalar-pseudoscalar interactions}
 \label{sec2a}

\begin{figure}[htpb]
 	\includegraphics[scale=0.45]{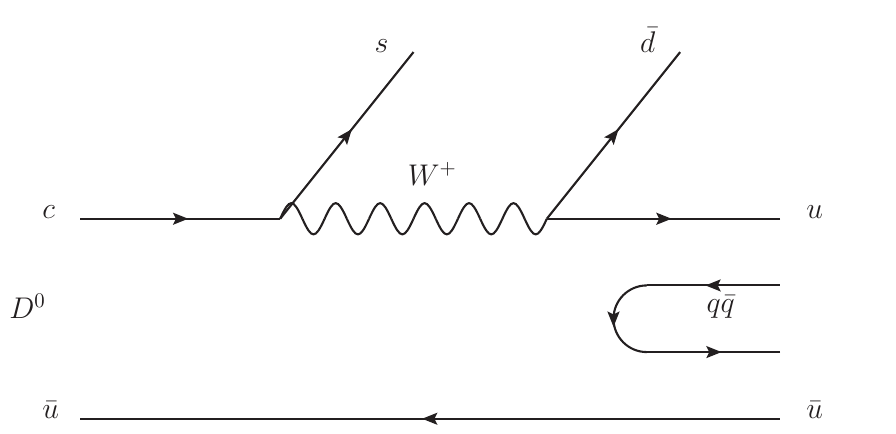}
 	\caption{Feynman diagram of $D^0\to \bar{K}^0u\bar{u}$ and hadronization of  the $u\bar{u}$ through $q\bar{q}$ with vacuum quantum numbers.}
 	\label{fig.quarklevel}
 	\end{figure}

Analogous to Refs.~\cite{Xie:2016evi,Miyahara:2015cja,Duan:2020vye}, the Cabibbo-favored process $D^0\to \bar{K}^0\pi^0\pi^0$ could be divided into three steps: the weak decay, the hadronization, and the final state interactions. For the first step, the $c$ quark of the initial state $D^0$ meson weakly decays into an $s$ quark and a $W^+$ boson, followed by the $W^+$ boson decaying into $\bar{d}u$ quarks. Next, as depicted in Fig.~\ref{fig.quarklevel}, the $s$ quark from the $c$ decay, and the $\bar{d}$ quark from the $W^+$ decay, hadronize into the $\bar{K}^0$ meson, while the $u$ quark from the $W^+$ decay and the $\bar{u}$ of the initial $D^0$, together with the antiquark-quark pair $\bar{q}{q}(=\bar{u}u+\bar{d}d+\bar{s}s)$ created from the vacuum with the quantum numbers $J^{PC}=0^{++}$, hadronize into two mesons as follows,
\begin{eqnarray}
 u(\bar{u}u+\bar{d}d+\bar{s}s)\bar{u}
 =\sum_{i=1}^3 M_{1i}×M_{i1}=\left(M^2\right)_{11},  \label{eq:hadronpair}  
\end{eqnarray}
where $i=1$, $2$, $3$ represent $u$, $d$, $s$ quarks. The $M$ is the $q\bar{q}$ matrix,
 	\begin{eqnarray}
 		M&=&\begin{pmatrix}
 			u\bar{u} & u\bar{d} & u\bar{s}\\
 			d\bar{u} & d\bar{d} & d\bar{s}\\
 			s\bar{u} & s\bar{d} & s\bar{s}\\
 		\end{pmatrix}	,
 		\label{eq.M}
 	\end{eqnarray}
which could also be expressed with the pseudoscalar meson basis as follows~\cite{Dai:2021owu,Molina:2019udw}, 	
 \begin{eqnarray}
 		M=
 		\begin{pmatrix}
 			\dfrac{\pi^0}{\sqrt{2}}+\dfrac{\eta}{\sqrt{3}}+\dfrac{\eta'}{\sqrt{6}}
 			&\pi^+  & K^+\\
 			\pi^-   &-\dfrac{\pi^0}{\sqrt{2}}+\dfrac{\eta}{\sqrt{3}}+\dfrac{\eta'}{\sqrt{6}}  & K^0\\
 			K^- & \bar{K}^0 & -\dfrac{\eta}{\sqrt{3}}+\dfrac{2\eta'}{\sqrt{6}}  \\
 		\end{pmatrix},
 		\end{eqnarray}
   where we adopt the approximate $\eta-\eta'$ mixing~\cite{Lyu:2023ppb,Bramon:1992kr}\footnote{{According to Review of Particle Physics (PRR)~\cite{ParticleDataGroup:2022pth}, the mixing angle is between $-10^\circ$ and $-20^\circ$, and a  recent Lattice calculations support the value $\theta_P=-14.1^\circ \pm 2.8^\circ$ by reproducing the masses of the $\eta$ and $\eta'$~\cite{Christ:2010dd}.
In this work, we adopt the mixing from Ref.~\cite{Bramon:1992kr}, as follows,
\begin{eqnarray}
\eta\sim \frac{1}{\sqrt{3}}(u\bar{u}+d\bar{d}-s\bar{s}),\nonumber \\
\eta'\sim \frac{1}{\sqrt{6}}(u\bar{u}+d\bar{d}+2s\bar{s}).\nonumber
\end{eqnarray}.}}.  
Then all the possible meson-meson components of Eq.~(\ref{eq:hadronpair}) are,
\begin{eqnarray}
\left(M^2\right)_{11}  &=&\dfrac{1}{2}\pi^0\pi^0+\dfrac{1}{3}\eta\eta+\dfrac{2}{\sqrt{6}}\pi^0\eta  \nonumber\\
 &&+\pi^+\pi^-+ K^+K^- , \label{eq:hadroncomponment}
\end{eqnarray}
where we have omitted possible components involving the $\eta'$  because it has a large mass and does not play a role in the generation of $f_0(980)$. Meanwhile, since the $\pi^0\eta$ system has the isospin $I=1$, and does not couple to $f_0(500)$ and $f_0(980)$, we will do not consider its contribution.

\begin{figure}[htbp]
 			\center
 		
 		\subfigure[]{		
 	\includegraphics[scale=0.4]{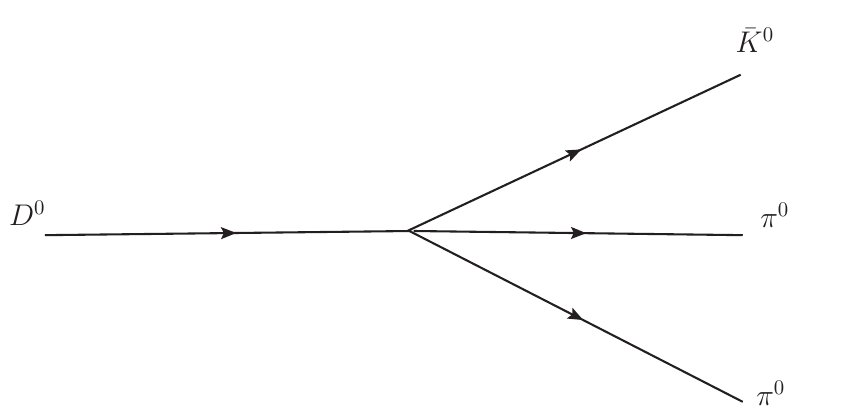}
 	}
 			
 	\subfigure[]{	
 		\includegraphics[scale=0.4]{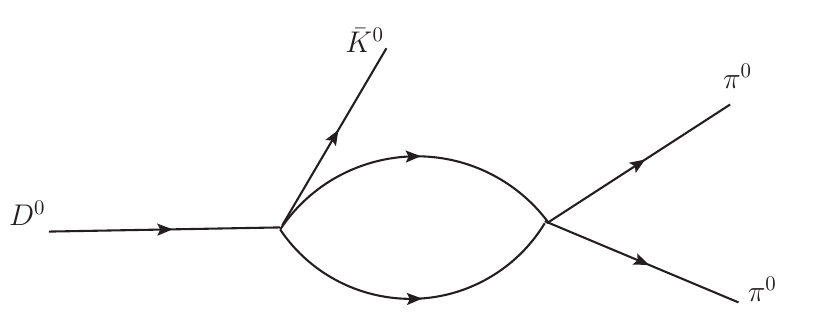}
    }	
    \caption{The tree diagram (a) and the  $S$-wave pseudoscalar meson-pseudoscalar meson interactions (b) for the process $D^0\to \bar{K}^0\pi^0\pi^0$.}	\label{fig:mechanism}
 	\end{figure}

As we discussed above, the Cabibbo-favored process $D^0\to \pi^0\pi^0\bar{K}^0$ could  happen via the tree diagram of Fig.~\ref{fig:mechanism}(a), and the final state interaction of the $S$-wave pseudoscalar meson-pseudoscalar meson of Fig.~\ref{fig:mechanism}(b).
 Then, the amplitude for the Figs.~\ref{fig:mechanism}(a) and \ref{fig:mechanism}(b) could be expressed as,
 \begin{eqnarray}
 	\mathcal{T}^{S}&=&V_p \left( h_{\pi^0\pi^0}+h_{\pi^0\pi^0}\times\frac{1}{2}G_{\pi^0\pi^0}t_{\pi^0\pi^0\to \pi^0\pi^0}\right.\nonumber \\
  && +h_{\eta\eta}\times\frac{1}{2}\times G_{\eta \eta}t_{\eta \eta \to \pi^0\pi^0}\nonumber\\
 	&&	+h_{\pi^+\pi^-}\times G_{\pi^+\pi^-} t_{\pi^+\pi^-\to\pi^0\pi^0} \nonumber \\
  && + \left.h_{K^+K^-}\times G_{K^+K^{-} } t_{K^+K^-\to \pi^0\pi^0} \right.) , \label{eq:ampswave}
\end{eqnarray}
where  the $V_p$ stands for the strength of the weak production vertex and could be treated as a constant, and
the coefficients $h_i$ represent the production weights of the corresponding components in the preliminary decay, which could be obtained from Eq.~(\ref{eq:hadroncomponment}),
\begin{eqnarray}
    h_{\pi^0\pi^0}=\frac{1}{2},~h_{\eta\eta}=\frac{1}{3},~
    h_{\pi^+\pi^-}=1,~h_{K^+K^-}=1.
\end{eqnarray}
 Here we introduce the extra factor ${1}/{2}$ for the loop systems of $\pi^0\pi^0  $ and $ \eta \eta $ with identical  particles, as done in Ref.~\cite{Xie:2014tma}.
 The $G_l$ is the loop function of two-meson propagator, and the $t_{i\to j}$ is the transition amplitude between the coupled-channels~\cite{Oller:1997ti},
 The $t$-matrix can be obtained by solving the Bethe-Salpeter equation~\cite{Oller:1997ti},
 \begin{equation}
 		T=[1-VG]^{-1}V, \label{eq:BS}
 	\end{equation} 
  where $V$ is a $5\times 5$ matrix of the interaction kernel for five channels $\pi^+\pi^-$, $\pi^0\pi^0$, $K^+K^-$, 
  $K^0\bar{K}^0$, and $\eta \eta $ with isospin $I=0$, and the 
  explicit expressions of the $5\times 5$ matrix elements are 
  given by~\cite{Liang:2014tia,Dias:2016gou}, 	
\begin{align}
& V_{11}=-\dfrac{1}{2f^2}s,
 V_{12}=-\dfrac{1}{\sqrt{2}f^2}(s-m^2_\pi),\nonumber\\
& V_{13}=-\dfrac{1}{4f^2}s,
 V_{14}=-\dfrac{1}{4f^2}s,\nonumber\\
& V_{15}=-\dfrac{1}{3\sqrt{2}f^2}m^2_{\pi},
 V_{22}=-\dfrac{1}{2f^2}m^2_{\pi},\nonumber\\
 &V_{23}=-\dfrac{1}{4\sqrt{2}f^2}s,
 V_{24}=-\dfrac{1}{4\sqrt{2}f^2}s,\nonumber\\
 &V_{25}=-\dfrac{1}{6f^2}m^2_{\pi},
 V_{33}=-\dfrac{1}{2f^2}s, 
 V_{34}=-\dfrac{1}{4f^2}s,\nonumber\\
 &V_{35}=-\dfrac{1}{12\sqrt{2}f^2}(9s-6m^2_{\eta}-2m^2_{\pi}),\nonumber\\
& V_{44}=-\dfrac{1}{2f^2}s,  
 V_{45}=-\dfrac{1}{12\sqrt{2}f^2}(9s-6m^2_{\eta}-2m^2_{\pi}), \nonumber\\
 &V_{55}=-\dfrac{1}{18f^2}(16m^2_K-7m^2-{\pi}),
\end{align}		
 with the pion decay constant $f=93$~MeV and the  invariant mass square $s$ of the meson-meson system. The masses of the mesons are taken from RPP~\cite{ParticleDataGroup:2022pth}. We have take the unitary normalization $\left| \eta\eta\right>\to \dfrac{1}{\sqrt{2}}\left| \eta\eta \right>$,
 	$\left| \pi^0\pi^0\right> \to \dfrac{1}{\sqrt{2}}\left|
  \pi^0\pi^0\right>$ to account for the identity of particles 
  when using the $G$ function without an extra factor in 
  Eq.~(\ref{eq:BS})~\cite{Liang:2014tia}. The loop function for two-meson system is,
  \begin{equation}
 		G_l = i\int\dfrac{d^4q}{(2\pi)^4}\dfrac{1}{(P-q)^2-m^2_1+i\epsilon}\dfrac{1}{q^2-m^2_2+i\epsilon},
   \label{eq.10}
 	\end{equation}
 where the $m_1$, $m_2$ are the masses of mesons in the $l$-channel, $q$ is the four-momentum of one meson, and the $P$ is the total four-momentum, with $s=P^2$. Since the loop function $G$ in Eq.~(\ref{eq.10}) is logarithmically divergent~\cite{Oller:1998hw}, we have two methods to solve this singular integral, either using the three-momentum cut-off method, or the dimensional regularization method, {by introducing free parameters, which are determined by giving rise to the pole position of the resonance.} In this work,  the integral of the loop function is performed integrating exactly for $q^0$ and implementing a cut off $q_{\rm max}$  for the three-momentum~\cite{Oller:1997ti}, 	
 \begin{equation}
G_l=\int_{0}^{q_{\rm max}}\dfrac{|\vec{q}\,|^2d|\vec{q}|}{(2\pi)^2}\dfrac{\omega_1+\omega_2}{\omega_1\omega_2[s-(\omega_1+\omega_2)^2+i\epsilon]},
 	\end{equation}
where the meson energies are $\omega_i=\sqrt{\vec{q}^{\,2}+m_i^2}$ with the subindex $i$ standing for the two intermediate mesons in the $l$-channel.
It should be stressed that, in Ref.~\cite{Oller:1997ti}, only $\pi\pi $ , $KK$ channel were considered, while in this work the channel $\eta\eta $ is also considered, since the threshold of this channel is not so much far away from the mass of  $f_0(980)$. The results obtained with or without the  $\eta\eta $ channel are very similar but the cut off needed with two channel is $q_{\rm max}$=903~MeV, while it is 600~MeV for three channels~\cite{Liang:2014tia}. The effective inclusion of new channels by means of change in the cutoff is common in many works~\cite{Liang:2014tia,Wang:2021naf}.  
  
In addition, the most reliable prediction range of the $S$-wave pseudoscalar meson-pseudoscalar meson interaction within the chiral unitary approach is up to $1100\sim1200$~MeV, and one can not use this model for higher invariant masses. 
In order to present a reasonable $\pi^0\pi^0$  invariant mass distribution, we evaluate the $Gt$ combinations of Eq.~(\ref{eq:ampswave}) up to $M_{\rm cut}=1.1$~GeV, and from there on we multiply $Gt$ by a smooth factor to make it gradually decreasing at large $M_{\pi^0\pi^0}$, as done in Ref.~\cite{Debastiani:2016ayp},
 \begin{equation}
Gt(M_{\pi^0\pi^0})=Gt(M_{\rm cut})e^{-\alpha (M_{\pi^0\pi^0}-M_{\rm cut})},~~  M_{\pi^0\pi^0}>M_{\rm cut},
 \end{equation} 
 Here, we consider $\alpha$=0.0054~MeV, which is the smoothing factor~\cite{Debastiani:2016ayp}.

 \subsection{The contribution from  $\bar{K}^{*}{(892)}$} 
   \label{sec.2b}
 	
\begin{figure}[htpb]
 		\centering
 		\includegraphics[scale=0.5]{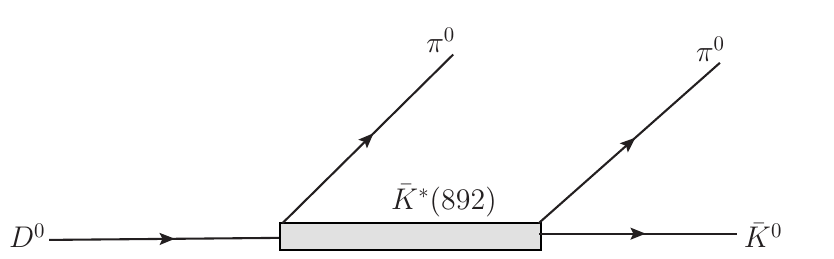}
 		\caption{The decay $D^0\to \pi^0\bar{K}^{*}(892)^0\to \pi^0 \pi^0\bar{K}^0$ via the intermediate vector $ \bar{K}^{*}(892)^0$.}
 		\label{fig.3}
 	\end{figure}

 	In addition to the contribution from the $S$~wave pseudoscalar meson-pseudoscalar meson interactions, we have also considered the contribution from intermediate vector meson $\bar{K}^{*}(892)$, as depicted in Fig.~\ref{fig.3}	
 	
 Since the vector meson  $\bar{K}^{*}(892)^0$ decays into $\pi^0\bar{K}^0$ in $P$-wave, the corresponding Lagrangian for the vertex of $VPP$ is given as,
\begin{equation}  
 		\mathcal{L}=-ig\langle [P, \partial_\mu P] V^\mu \rangle ,
 	\end{equation}
where the $g={M_V}/ {2f}$ is the effective coupling, and the $P$ and $V$ represents the matrix of the pseudoscalar meosn and vector meson, respectively. The matrix of pseudoscalar mesons is given by Eq.~(\ref{eq.M}), and the matrix of  the vector mesons $V$ is written as~\cite{Zhu:2022wzk},	
\begin{eqnarray}
 		V=
 		\begin{pmatrix}
 			\dfrac{\rho^0}{\sqrt{2}}+\dfrac{\omega}{\sqrt{2}} & \rho^+ & K^{*}\\
 			\rho^- &  -\dfrac{\rho^0}{\sqrt{2}}+\dfrac{\omega}{\sqrt{2}} &  K^{*0}\\
 			K^{*-} &\bar{K}^{*0} & \phi \\
 		\end{pmatrix}.
 	\end{eqnarray}
 	Thus, the potential of the  $\bar{K}^{*}(892)^0\to \pi^0\bar{K}^0$ (we denote this $\pi$ as the first $\pi_1$) is,
\begin{equation}
V_{\bar{K}^{*0}\to \pi^0\bar{K}^0}= \langle i \mathcal{L}\rangle 
 		=\frac{g}{\sqrt{2}}(p_{\bar{K}^0}-p_{\pi_1^0})_\mu\epsilon^\mu _{\bar{K}^{*0}}.
 	\end{equation} 
 	
 	For the vertex of $D^0\to \pi^0 \bar{K}^{*0}$ (we denote this $\pi$ as the second $\pi_2$), we could have the potential with the same structure,
\begin{equation}
 		V_{D^0\to \pi^0\bar{K}^{*}}=g_{D^0\pi^0\bar{K}^{*0}}\epsilon^\nu_{\bar{K}^{*0}} (p_{D^0}+p_{\pi_2^0})_\nu,
 	\end{equation}
 	where the $p_{D^0}$ and $p_{\pi^0_2}$ stand for $D^0$ and $\pi^0$ four-momenta respectively, and the $g_{D^0\pi^0\bar{K}^{*0}}$ is an effective coupling for the vertex of $D^0\to\pi^0\bar{K}^{*0}$. Then the amplitude for the  $D^0\to \pi^0 \bar{K}^{*}(892)\to \pi^0\pi^0\bar{K}^0$  could be written as~\cite{Zhu:2022wzk,Zhu:2022guw},
 	\begin{eqnarray}
\mathcal{T}^{\bar{K}^{*}}&=&\dfrac{g_{D^0 \pi^0\bar{K}^{*0}} g}{\sqrt{2}} \dfrac{1}{M_{\pi_1^0\bar{K}^0}^2-m^2_{\bar{K}^{*}}+iM_{\bar{K}^{*}}\Gamma_{\bar{K}^{*}}}\nonumber\\	&&\left[-\left(m^2_{\bar{K}^0}-m^2_{\pi^0_1}\right)\left(1-\dfrac{M^2_{\pi^0_1\bar{K}^0}}{m^2_{\bar{K}^{* }}}\right)\right. \nonumber \\
&& + 2p_{\pi^0_1}p_{\pi^0_2}\dfrac{m^2_{\bar{K}^{*}}+m^2_{\bar{K}^0}-m^2_{\pi^0_1}}{m^2_{\bar{K}^{*}}}\nonumber\\
 &&\left. +2p_{\bar{K}^0}p_{\pi^0_2}\dfrac{m^2_{\bar{K}^0}-m^2_{\bar{K}^{*}}-m^2_{\pi^0_1}}{m^2_{\bar{K}^{*}}}\right]\nonumber \\
 &&+ {\rm exchange} \left(\pi^0_1 \leftrightarrow \pi^0_2\right),
 	\end{eqnarray}
  with
  \begin{align}
 		&M^2_{\pi^0_1\bar{K}^0}=(p_{\bar{K}^0}+p_{\pi^0_1})^2\\
 		&p_{\pi^0_1}\cdot p_{\pi^0_2}=\frac{M^2_{\pi^0_1\pi^0_2}-2m^2_{\pi^0}}{2}\nonumber\\
 		&p_{\bar{K}^0}\cdot p_{\pi^0_2}=\frac{m^2_{D^0}+m^2_{\pi^0_1}-M^2_{\pi^0_1\pi^0_2}-M^2_{\pi^0_1 \bar{K}^0}}{2},
 	\end{align}
where the exchange $\left(\pi^0_1 \leftrightarrow \pi^0_2\right)$ stands for replacement the particle $\pi^0_1$  of the term  with particle $\pi^0_2$. 
With the above formalism, the double differential decay width can be expressed as, 	
\begin{equation}
 		\dfrac{d^2\Gamma}{dM_{\pi^0_1\pi^0_2}dM_{\pi^0_1\bar{K}^0}}
 		= \dfrac{M_{\pi^0_1\pi^0_2} M_{\pi^0_1\bar{K}^0}}{128\pi^3m^3_{D^0}} 
 		\left(\left|\mathcal{T}^S\right|^2+\left|\mathcal{M}^{\bar{K}^{*}}\right|^2\right).
 		\label{eq:dwidth}
\end{equation}
One could obtain the invariant mass distributions $d\Gamma/dM_{\pi^0_1\pi^0_2}$ and $d\Gamma/dM_{\pi^0_1\bar{K}^0}$ by integrating Eq.~(\ref{eq:dwidth}) over each of the invariant mass variables. For a given value of $M_{12}$, the range of $M_{23}$ is determined by~\cite{ParticleDataGroup:2022pth},
\begin{align}
	&\left(m_{23}^2\right)_{\min}=\left(E_2^*+E_3^*\right)^2-\left(\sqrt{E_2^{* 2}-m_2^2}+\sqrt{E_3^{* 2}-m_3^2}\right)^2, \nonumber\\
	&\left(m_{23}^2\right)_{\max}=\left(E_2^*+E_3^*\right)^2-\left(\sqrt{E_2^{* 2}-m_2^2}-\sqrt{E_3^{* 2}-m_3^2}\right)^2, 
\end{align}
where $E_2^{*}$ and $E_3^{*}$ are the energies of particles 2 and 3 in the $M_{12}$ rest frame, which are written as,
\begin{align}
	&E_2^{*}=\dfrac{M_{12}^2-m_1^2+m_2^2}{2M_{12}}, \nonumber\\
	&E_3^{*}=\dfrac{M_{D^0}^2-M_{12}^2-m_3^2}{2M_{12}},
\end{align}
with $m_1$, $m_2$, and $m_3$ are the masses of involved particles 1, 2, and 3, respectively. All the masses and widths of the particles involved in this work are taken from the RPP \cite{ParticleDataGroup:2022pth}.

 \section{ Results and discussions}
 \label{sec.3}

In our formalism discussed above, there are two parameters, the normalization factor $V_p$ of Eq.~(\ref{eq:ampswave}), and the effective couplings $g_{D^0 \pi^0\bar{K}^{*}} $. Since the experimental measurements of the CLEO are the $\pi^0\pi^0$ and $\pi^0K_S^0$ invariant mass square distributions, we could introduce a parameter $\beta=g_{D^0 \pi^0\bar{K}^{*0}} g/V_p$ to account for the relative weight of the contribution from $\bar{K}^{*0}$ with respective to the one from the $S$-wave pseudoscalar meson-pseudoscalar meson interactions. In addition, the bin sizes of the $\pi^0\pi^0$ and $\pi^0K_S^0$ invariant mass square distributions are 0.038~GeV$^2$ and 0.055~GeV$^2$, respectively. Thus, in order to account for the different bin sizes of  the $\pi^0\pi^0$ and $\pi^0K_S^0$ invariant mass square distributions simultaneously, we need to adopt two different normalization factor $V_p$ and $V'_p$ for the $\pi^0K_S^0$  and $\pi^0\pi^0$ invariant mass square distributions, and has the relationship $(V'_p)^2=\frac{1}{2}\times \frac{38}{55}(V_p)^2$, where we introduce the factor $\frac{1}{2}$ because there are two $\pi^0$ in the final states.

 First, we have performed a fit to the experimental data of the $\pi^0\pi^0$ and $\pi^0 K^0$ invariant mass distributions~\cite{CLEO:2011cnt}. Then we obtain $\chi^2/d.o.f.$ is 237.8/(47 + 48-2) = 2.56, and the fit parameters $(V'_p)^2=(6.275\pm0.2472)\times10^{9}$, $\beta=(9.5195\pm 0.313)\times10^{-2}$. 
 	
\begin{figure*}[htbp]
 
 \subfigure[]{
 \includegraphics[scale=0.7]{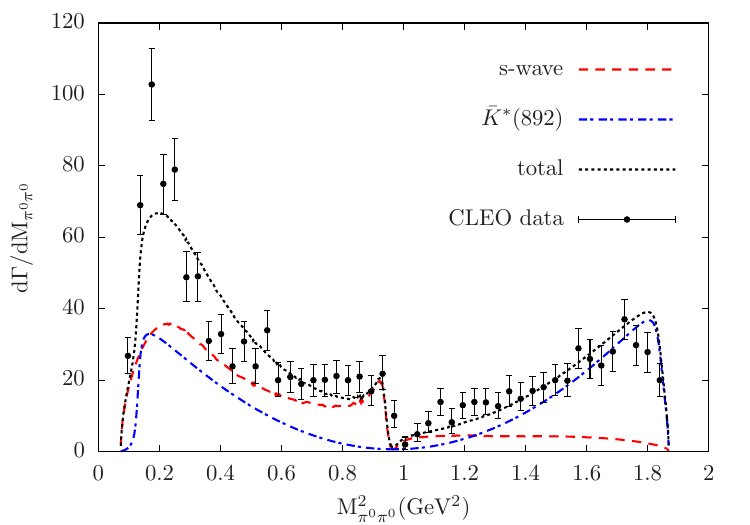}
 		}
 \subfigure[]{
\includegraphics[scale=0.7]{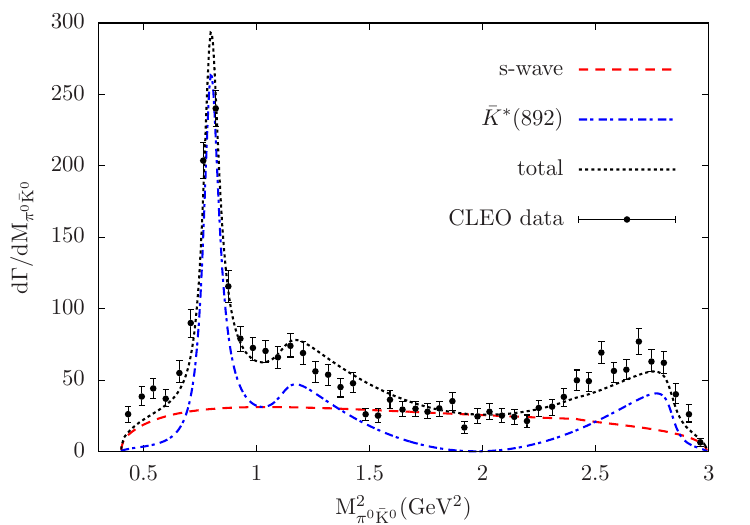}}
 
 \caption{The $\pi^0\pi^0$ (a) and $\pi^0\bar{K}^0$ (b) invariant mass distributions for the process $D^0\to \pi^0\pi^0\bar{K}^0$. The blue dashed curves and red dotted curves correspond to the contributions from the $S$-wave pseudoscalar meson-pseudoscalar interactions and the vector meson $\bar{K}^{*0}$, respectively, the solid black curves stand for the results of the full model. In addition, The experimental data of CLEO is represented by the points with error bars~\cite{CLEO:2011cnt}.   }
 		\label{fig.jieguo}
 	\end{figure*}

 	Next, we have calculated the $\pi^0\pi^0$ and $\pi^0\bar{K}^0$ invariant mass distribution with the fitted parameters in Fig.~\ref{fig.jieguo}. The black points with error bars labeled as `CLEO data' are the CLEO-c data taken from Ref.~\cite{CLEO:2011cnt}. One can find that, in the $\pi^0\pi^0$ invariant mass distribution, our results could well reproduce the structure of the cusp followed by a dip,  which could be associated with the scalar meson $f_0(980)$, dynamically generated from the $S$-wave pseudoscalar meson pseudoscalar meson interactions. Meanwhile, our results show that the peak around $M^2_{\pi^0\pi^0}=0.2$~GeV$^2$ is mainly due to the scalar meson $f_0(500)$ and the reflection of the vector meson $\bar{K}^{*0}$. {Within the coupled-channel Bethe-Salpeter approach with the meson-meson potentials provided by the lowest order chiral lagrangian~\cite{Oller:1997ti}, one can simultaneously find a broad resonance $f_0(500)$, which couples strongly to $\pi\pi$ channel, and the one $f_0(980)$, which couples strongly to $K\bar{K}$ channel, which implies that these two states have a different dynamically internal structure.}

  In addition, our results for the $\pi\bar{K}^0$ invariant mass distribution agree well with the CLEO measurements.  The narrow peak structure around $M^2_{\pi^0\bar{K}^0}=0.7$~GeV$^2$ should be associated with the vector meson $K^{*}{(892)}$, while the contribution from the $S$-wave pseudoscalar meson-pseudoscalar meson interaction serves as a smooth background in the $\pi\bar{K}^0$ invariant mass distribution.  

\begin{figure}[htbp]\center {\includegraphics[scale=0.9]{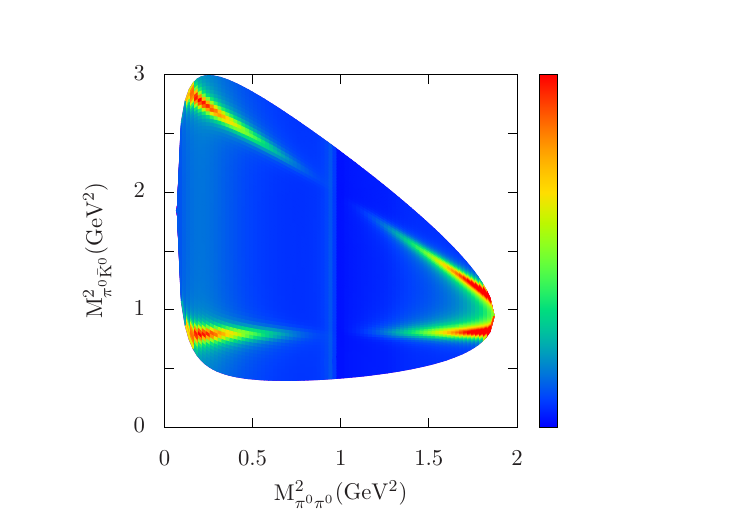} 	}
 	
 	\caption{The Dalitz plot of '$M^2_{\pi^0\pi^0}$' vs.'$M^2_{\pi^0\bar{K}^0}$' for the process $D^0\to \pi^0\pi^0\bar{K}^0$. }
 	\label{fig.dalitz}
 \end{figure}

 	 We also present the Dalitz plot of `${M^2_{\pi^0\bar{K}^0}}$' and `${M^2_{\pi^0\pi^0}}$' in Fig.~\ref{fig.dalitz}, which clearly shows the structures of $f_0(980)$ and $\bar{K}^{*}(892)$. One can also find that the peak near the  $\pi^0\pi^0$ threshold is mainly due to the scalar meson $f_0(500)$ and the reflection of the vector meson $\bar{K}^{*0}$.

   Finally, with the fitted parameters, we could estimate that the $S$-wave $\pi^0\pi^0$ interaction contributes about $47.7\%$ of the total decay rate, larger than the experimental data $(28.9\pm6.3\pm 3.1)\%$, while the intermediate resonance $K^*(892)$ contributes about $52.3\%$ of the total decay rate, consistent with the experimental data $(65.6\pm 5.3\pm 2.5)\%$ within the uncertainties~\cite{CLEO:2011cnt}. It should be pointed out that the contribution from the scalar $f_0(500)$ was not taken into account by CLEO~\cite{CLEO:2011cnt}, which plays an important role in the near-threshold enhancement of the $\pi^0\pi^0$ invariant mass distribution.

 	\section{SUMMARY AND CONCLUSION}
 	\label{sec.4}
 	
Motivated by the CLEO measurements on the process $D^0\to \pi^0\pi^0\bar{K}^0$~\cite{CLEO:2011cnt}, we have investigated this process by taking into account the contribution from the $S$-wave pseudoscalar meson-pseudoscalar meson interaction within the chiral unitary method, which could dynamically generate intermediate resonance states $f_0(980)$ and $f_0(500)$, and also the contribution from the intermediate vector meson $\bar{K}^*{(892)}$. 

First, we have performed a fit to the experimental data of the $\pi^0\pi^0$ and $\pi^0K_S^0$ invariant mass distributions. With the fitted parameters, our results could well reproduce the cusp structure followed a dip around 1~GeV in the $\pi^0\pi^0$ invariant mass distribution, which is mainly due to the contribution of scalar $f_0(980)$. Our results indicate that the peak near the $\pi^0\pi^0$ threshold is due to the scalar meson $f_0(500)$ and the reflection of the $\bar{K}^*$.
Our results of the $\pi^0\bar{K}^0$ invariant mass distribution are in good agreement with the CLEO data, which supports that the peak around 892 MeV in the $\pi^0\bar{K}^0$ invariant mass distribution should be associated with $\bar{K}^*{(892)}$. 

{As we discussed in the introduction, there have different explanations for the nature of the scalar mesons $f_0(500)$ and $f_0(980)$. Our work, taking into account $S$-wave pseudoscalar meson-pseudoscalar meson interaction within the chiral unitary method, could well describe the CLEO measurements of the process $D^0\to \pi^0\pi^0\bar{K}^0$, especially the cusp structure followed a dip around 1~GeV in the $\pi^0\pi^0$ invariant mass distribution, which favors the molecular explanation of the scalar mesons $f_0(500)$ and $f_0(980)$.}

\section*{Acknowledgments}

This work is partly supported by  the National Key R\&D Program of China under Grand No. 2024YFE0105200, the Natural Science Foundation of Henan under Grant No. 232300421140 and No. 222300420554, the National Natural Science Foundation of China under Grant No. 12475086 and No. 12192263.

 \end{document}